\documentclass[sigconf]{acmart}

\AtBeginDocument{%
  }

\setcopyright{acmlicensed}
\copyrightyear{2026}
\acmYear{2026}
\acmDOI{XXXXXXX.XXXXXXX}
\acmConference[SIGCSE '27]{58th ACM Technical Symposium on Computer Science
Education V.1 (SIGCSE TS 2027)}{February 17--20,
  2027}{Sacramento, CA}

\acmISBN{978-1-4503-XXXX-X/2018/06}

\usepackage{colortbl}
\usepackage{graphicx}
\usepackage{tabularx}
\usepackage{float}
\usepackage{array}
\usepackage[table]{xcolor}

\usepackage{booktabs,tabularx,multirow,xcolor,colortbl}
\usepackage{footnote}\makesavenoteenv{tabularx}  
\newcolumntype{A}{>{\raggedright\arraybackslash\hsize=1.02\hsize}X}
\newcolumntype{B}{>{\raggedright\arraybackslash\hsize=1.16\hsize}X}
\newcolumntype{C}{>{\raggedright\arraybackslash\hsize=0.82\hsize}X}

\usepackage{multirow}
\usepackage{makecell}
\usepackage{tablefootnote}
\usepackage{enumitem}
\usepackage{color}

\usepackage[most]{tcolorbox}
\usepackage{xcolor}

\newtcolorbox{instructorbox}{
  colback=white,
  colframe=black,
  boxrule=0.5pt,
  arc=1pt,
  boxsep=2pt,
  left=3pt,
  right=3pt,
  top=2pt,
  bottom=2pt,
  toptitle=1pt,
  bottomtitle=1pt,
  before skip=3pt,
  after skip=3pt,
  fonttitle=\bfseries\small,
  title=Instructor Implication
}

\newtcolorbox{designerbox}{
  colback=white,
  colframe=black,
  boxrule=0.5pt,
  arc=1pt,
  boxsep=2pt,
  left=3pt,
  right=3pt,
  top=2pt,
  bottom=2pt,
  toptitle=1pt,
  bottomtitle=1pt,
  before skip=3pt,
  after skip=3pt,
  fonttitle=\bfseries\small,
  title=Tool Designer Implication
}

\usepackage{booktabs}
\usepackage{pifont}
\usepackage{soul}
\usepackage{xcolor}

\definecolor{planhighlight}{gray}{0.92}

\newcommand{\planstep}[1]{%
  \begingroup
  \sethlcolor{planhighlight}%
  \hl{#1}%
  \endgroup
}

\begin{document}



\title{A Plan-Tracing Interface for AI-Supported Algorithm Planning}
\author{Yoshee Jain}
\email{yosheej2@illinois.edu}
\affiliation{%
  \institution{University of Illinois Urbana-Champaign}
  \city{Urbana}
  \state{IL}
  \country{USA}
}

\author{Heejin Do}
\email{heejin.do@ai.ethz.ch}
\affiliation{%
  \institution{ETH Zurich}
  \country{Switzerland}
}

\author{Zihan Wu}
\email{zihan.wu@maine.edu}
\affiliation{%
  \institution{University of Maine}
  \city{Orono, ME}
  \country{USA}
}

\author{April Yi Wang}
\email{april.wang@inf.ethz.ch}
\affiliation{%
  \institution{ETH Zurich}
  \country{Switzerland}
}
\renewcommand{\shortauthors}{Anonymous et al.}


\begin{abstract}

\noindent \textbf{Background:} Planning an algorithm in natural language allows learners to get formative feedback on their approach before coding. But these descriptions can be ambiguous, making it challenging for learners to translate them into code and for LLMs to provide feedback.
\textbf{Objective:} To address these challenges, we introduce \textit{plan tracing}, in which learners manually simulate how the described algorithm would execute on a concrete input. We develop an interface that enables plan tracing and allows learners to receive AI feedback on their plans before writing code.
\textbf{Method:} We report on an exploratory between-subjects study with 20 participants who solved an algorithm design task, with or without the plan tracing interface. We observed how plan tracing shaped students' plans, the feedback they received, and their experiences using the interface.
\textbf{Findings:} Students who performed plan tracing wrote plans with fewer code-like steps but more goal-driven descriptions. We did not detect a difference in the quality of the LLM feedback between conditions. Students used plan tracing to debug and verify their strategy, describing it as tedious but worthwhile when they were uncertain about their solution.
\textbf{Implications:} We reflect on the design and use of our tool, identifying what worked, what didn't, and why, and offer recommendations for instructors and tool designers.

\end{abstract}



\maketitle

\section{Introduction}

\begin{figure*}[t]
    \centering
    \includegraphics[width=0.9\linewidth]{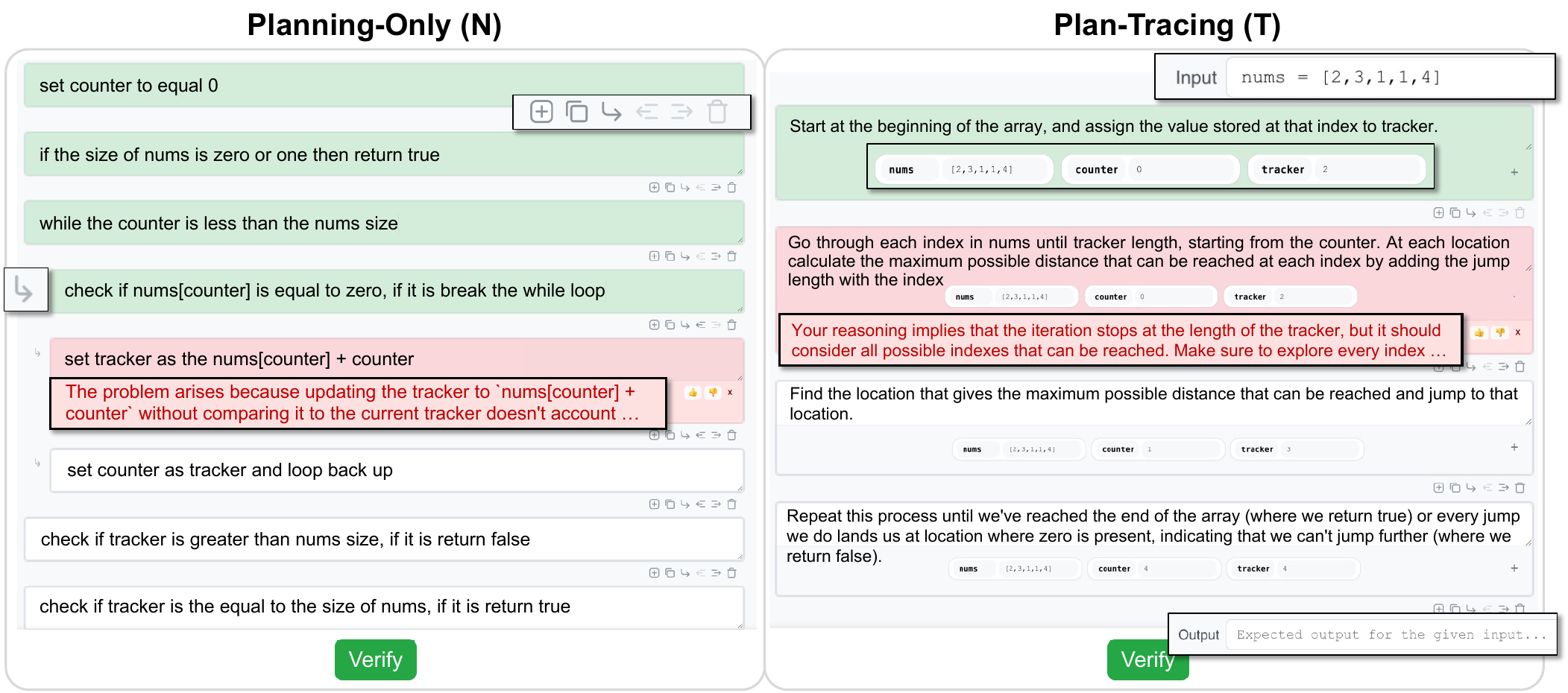}
    \caption{\textbf{Left:} Planning-only (N) interface, in which students decompose their solution into structured steps and optional sub-steps (indented) using textual explanations. \textbf{Right:} Plan-tracing interface (T), where students articulate textual explanations and additionally construct a step-by-step trace showing the progression of predefined variables on a given (but editable) input. In both conditions, learners iteratively refine their plan using feedback from LLMs prior to coding.}
    \label{fig:representative}
    
\end{figure*}

Programming requires learners to reason about high-level solution strategies and low-level implementation details, and novices often struggle to connect these two levels effectively~\cite{robins201912,castro2020qualitative}. \textit{\textbf{Planning}}, which is articulating an intended solution before writing code \cite{rivera2024iterative,rivera2024observations,cunningham_turing_tarpit}, is an extensively studied technique for managing this complexity by separating problem solving into two stages: what the program should do and how to implement it~\cite{soloway1986learning}. By making students' intended solution explicit before implementation, planning creates an opportunity for providing formative feedback on students’ underlying approach ~\cite{spohrer2013goal,rivera2024iterative}. Plans were traditionally evaluated manually, but because this is difficult to scale \cite{fowler2021should,burrows2015eras}, recent systems use LLMs to provide feedback. Some tools scaffold the planning \textit{process}, using LLMs to help learners decompose a problem and iteratively refine the natural-language representation of their solution~\cite{dbox}. Others \textit{evaluate} natural language artifacts by generating executable code from a description and testing it against instructor-defined test cases to return feedback~\cite{rivera2024iterative,smith2024code}. 

Both mechanisms, however, depend on an LLM correctly interpreting the student’s intent from their natural-language descriptions. This reliance on inference is especially challenging because unstructured natural-language plans may under-specify how an intended algorithm should behave, particularly when novice students use imprecise or non-standard descriptions~\cite{fowler2021should,mordechai2024novicode,chen2020validated,rivera2024observations}. LLMs may then misinterpret students' intent and provide incorrect or inconsistent feedback~\cite{babe-etal-2024-studenteval,dbox,rivera2024iterative}. This underspecification also affects learners since they need to resolve omitted details and ambiguous specifications when translating a plan into code~\cite{bonar2013preprogramming,pane2001studying}. Thus, for both learners and the LLMs, the key challenge is that a natural-language plan may convey the overall strategy but leaves its individual steps insufficiently concrete.

\textit{\textbf{Code tracing}}, that is manually simulating an algorithm’s stepwise execution on a concrete input, is a well-established technique for helping learners understand and debug program behavior~\cite{yang2025decoding,fitzgerald2008debugging,robobug,xie2018explicit,cunningham2017sketching}. We repurpose code tracing and introduce \textbf{\textit{plan tracing}} which is simulating how the algorithm described in a natural-language plan would execute on a concrete input (Figure~\ref{fig:representative}). We designed plan tracing to address the above-described challenges in two ways. For learners, tracing may reduce ambiguity by requiring them to specify how each plan step changes the program state on a concrete input, exposing missing conditions and underspecified operations. For the LLM, the trace may provide additional evidence of the intended behavior and reduce the amount it must infer from the natural-language plan when generating feedback.

In this paper, we present a plan-tracing interface that allows students to construct a natural-language plan, trace its execution on an input, and receive AI-generated feedback on their approach before writing code. We conducted a between-subjects study with 20 students to observe differences between plan tracing and natural-language planning alone. We asked the following questions:

\begin{description}[
style=multiline,
labelindent=0.6em,
leftmargin=4.2em,
labelwidth=2.8em,
labelsep=0.2em,
font=\normalfont\bfseries,
nosep
]
\item[RQ1] How does plan tracing influence how students construct and represent their plans?

\item[RQ2] What role does the trace play in shaping AI-generated feedback, and how do students experience that feedback?

\item[RQ3] How do students perceive and use plan tracing within their problem-solving process?

\end{description}

\label{sec:intro}

\section{Related Work}
\subsection{Planning and Tracing in CSEd}

To solve programming problems, learners must formulate a solution strategy, translate that strategy into code, and verify the resulting implementation~\cite{mccracken2001multi,robins2003learning}. Various scaffolds can support students in this challenging process, including subgoal-templates ~\cite{margulieux2012subgoal}, Parsons problems ~\cite {ericson2017solving}, code templates~\cite{cunningham_turing_tarpit,atkinsonfadedscaffolding}, and tracing strategies \cite{xie2018explicit}.

\subsubsection{Planning}
Within this landscape of the programming process, we focus on strategy formulation, also known as the \textit{planning} stage. Prior work has represented plans in multiple forms~\cite{rivera2024observations}, including unstructured natural-language descriptions~\cite{rivera2024iterative}, semi-structured subgoal-labeled templates~\cite{margulieux2012subgoal}, block-based artifacts~\cite{rivera2024observations}, programming plans~\cite{cunningham_turing_tarpit}, and pseudocode~\cite{saenz2022novices}. While more structured representations can provide scaffolding that benefits novices, more experienced learners can find them constraining or less authentic to programming~\cite{rivera2024observations,cunningham_turing_tarpit}. Unstructured descriptions, by contrast, are flexible and expressive, but may not clearly communicate students’ intent because students often use imprecise and non-standard terminology~\cite{fowler2021should,mordechai2024novicode,chen2020validated,rivera2024observations}. This may be detrimental for both learners who must resolve ambiguities when translating a plan into code~\cite{castro2020qualitative,bonar2013preprogramming} and automated systems that interpret natural language, whether translating into code~\cite{mordechai2024novicode, babe-etal-2024-studenteval} or grading it~\cite{smith2024code}.

\subsubsection{Tracing}
Code tracing is the process of stepping through a program's execution on a concrete input, and has been shown to help students understand and debug programs ~\cite{xie2018explicit, cunningham2017sketching, fitzgerald2008debugging}. By simulating how code executes line by line, learners use their notional machines to predict program behavior \cite{juha2013notional}, supporting \textit{program comprehension}~\cite{nelson2017comprehension}. Explicit strategies scaffold this process, such as writing down intermediate variable states, which improves novices' tracing accuracy \cite{xie2018explicit}. By following how a program’s actual behavior diverges from its intended behavior, learners can locate and fix errors, supporting \textit{debugging}~\cite{fitzgerald2008debugging,yang2025decoding}. Integrated development environments support this process with step-through debuggers ~\cite{stallman2002debugging, bennedsen2010bluej} that let learners inspect variable state as each line executes~\cite{hassan2024evaluating}. 

\subsubsection{Planning and Tracing}
A \textit{plan} captures a learner's intended strategy but often leaves operational details and conditions underspecified. Tracing can capture the \textit{concrete execution}, which can help surface the details a natural-language plan leaves implicit; however, it has traditionally been applied only to code. We bring these two ideas together by asking learners to trace the algorithm described by their natural-language plan before implementation, to see if tracing makes the plans' implicit behavior visible.

\subsection{Feedback on Plans}
Obtaining feedback on a proposed solution before it is implemented in code allows learners to refine their reasoning and approach early, leaving only translation to a later stage~\cite{spohrer2013goal,rivera2024iterative}. Traditionally, instructors and teaching staff reviewed students' plans and provided feedback \cite{de2009teaching}. However, grading open-ended, natural-language responses is time-consuming, inconsistent across graders, and delays feedback to learners \cite{fowler2021should,burrows2015eras}. Recent systems overcome this limitation by using LLMs for open-ended short-answer grading \cite{zhao2026ai}. Some tools \textit{help learners} decompose a problem and refine their natural-language solution~\cite{dbox}. Others \textit{evaluate} natural-language artifacts by prompting the LLM to generate code from descriptions and then testing that code against instructor-defined cases~\cite{rivera2024iterative, smith2024code}. While these approaches are promising for scalability, the feedback they produce can be unreliable. As novice learners often write imprecise and ambiguous descriptions, LLMs' inference of the intended behavior and missing details can be inaccurate. Prior work has found that these LLM assumptions may diverge from what the learner actually meant, leading to incorrect or inconsistent feedback~\cite{mordechai2024novicode, babe-etal-2024-studenteval}, thereby impeding learning. To address this issue, our study explores whether complementing natural-language plans with execution traces clarifies learners’ intent and supports the generation of more accurate and reliable LLM feedback.

\label{sec:background}

\section{Plan-Tracing Interface}
As shown in Figure~\ref{fig:representative}, we designed a planning environment in which students (1) decompose a problem into a natural-language plan, (2) trace how their plan executes on an input, (3) refine their plan using AI-generated feedback, and (4) implement their plan in code. 
We incorporated tracing for two purposes: (a) to help students examine and articulate their approach more precisely and (b) to provide the LLM with clear evidence of the algorithm's behavior.

\subsection{Designing the Plan-Tracing Workflow}
\subsubsection{Planning with Structured Natural Language.} 
\label{sec:system_planning}
Following prior work~\cite{dbox}, our interface allows students to represent their solution as an ordered list of steps, each described in natural language. 
Students can add steps and decompose them into finer-grained substeps, which are shown nested beneath their parent step. 
They can also edit, reorder, and remove steps as their approach develops. 
This hierarchical structure allows students to express the logic of their intended solution without committing to syntactic details.

\subsubsection{Tracing the Plan.} 
\label{sec:system_tracing}
The distinctive feature of our interface is that students trace the execution implied by their natural-language plan alongside its written steps. 
Students begin with a provided input and a predefined set of variables. 
For each plan step, they record how the values of these variables change as their described algorithm proceeds (Figure~\ref{fig:representative}, right). 
They may also modify the input to examine how their approach behaves on other cases. 
We designed this interaction to help students identify underspecified or incorrect steps while they were still developing their approach.

\subsubsection{Getting Feedback.} 
\label{sec:system_feedback}
After completing their plan and trace, students can request feedback. 
We delay access to feedback until a complete solution has been constructed so that the LLM informs students' revisions without directing their initial approach. 
During verification, the system sends OpenAI's GPT-4o the problem statement, the student's plan, the execution trace associated with each plan step, and a reference solution. 
The model identifies the first plan step that diverges from the reference approach. 
This \textit{reference-anchored feedback} is intended to help the model locate errors more reliably by comparing the student's plan with a known-correct solution rather than evaluating it in isolation. 
We also provide a fixed set of variables, rather than allowing students to define their own, so that the student and reference traces are in a comparable form. The feedback is designed to be \textit{fix-withholding} in that the LLM identifies the problematic step and explains why it is incorrect without revealing how to correct it. This design preserves the student's responsibility for revising the approach rather than allowing the LLM to give away the solution. The incorrect step is highlighted in red, with the feedback displayed alongside it. Students can revise and verify their plan iteratively in response to feedback.

\subsubsection{Implementing Plans in Code.}
\label{sec:system_coding}
We designed the workflow as a sequential process in which students first construct and verify a complete plan before implementing it in code. This ordering reflects an intentional separation between determining \textit{what} the algorithm should do and deciding \textit{how} to implement it in code~\cite{soloway1986learning}. The goal was for students to finalize the solution strategy during planning, allowing them to focus on implementation during coding.

\subsection{Exercise Authoring and Student Workflow}

Deploying plan tracing for new questions requires an instructor to specify a problem statement, a set of variables that students will trace, and a starter input on which the trace is initially performed. For the Jump Game task, we gave the variables \texttt{nums},  \texttt{counter} and \texttt{tracker} and provided the starter input \texttt{nums = [2,3,1,1,4]}. Finally, the instructor provides a reference solution, which the system uses to provide feedback on the correctness of the student's plan. Students see the problem, variables, and example input, write a natural-language plan and trace how the change in variables on the given input. The system compares the plan and trace against the instructor-provided reference and returns step-level feedback. Students revise and verify their approach iteratively before coding.

\label{sec:system}

\section{Evaluation}
\subsection{Method}
\subsubsection{Participants}
We recruited students with prior programming experience but limited proficiency with greedy algorithms from a large public university in the United States\footnote{A few participants were not affiliated with the university because recruitment materials were shared beyond university communication platforms by word of mouth.}, using advertisements (see table~\ref{tab:participants}).
Eligibility was assessed using participants' self-reported familiarity with greedy algorithms on a 5-point Likert scale. 
Students were randomly assigned to either the plan-tracing or the planning-only condition. 
After excluding one participant with high proficiency, our final sample consists of 20 participants, with 10 per condition. 
Each participant completed a 60-minute remote study session and received a \$25 USD Amazon gift card.

\vspace{-0.7em}
\begin{table}[H]
\centering
\small
\setlength{\tabcolsep}{4pt}
\caption{Participant demographics. N (planning-only condition); T (plan-tracing condition.}
\label{tab:participants}
\vspace{-10pt}
\scalebox{0.75}{
\begin{tabular}{l|cccc|ccc|ccc}
\hline
\textbf{Cond.} & 
\multicolumn{4}{c|}{\textbf{Academic level}} &
\multicolumn{3}{c|}{\textbf{Gender}} &
\multicolumn{3}{c}{\textbf{Coursework}} \\
\rowcolor[gray]{0.9}
 & Soph. & Junior & Senior & Grad & Male & Female & N/D & Intro & Data Struct. & Algo. \\
\hline
N & 4 & 4 & 1 & 1 & 7 & 2 & 1 & 9 & 8 & 7 \\
T & 3 & 2 & 0 & 5 & 7 & 3 & 0 & 10 & 10 & 7 \\
\hline
\end{tabular}
}
\vspace{-8pt}
\end{table}

\subsubsection{Study Environment and Conditions}
Participants joined the study remotely through Zoom. With their consent, we recorded their screens, microphones, and webcams. A web browser interface was made available so that participants could access the planning tools and complete the tasks detailed below. In the planning-only condition, participants described their proposed algorithm in natural language (Figure~\ref{fig:representative}, left). In the plan-tracing condition, participants were additionally required to complete a step-by-step execution trace of the given variables on an input (Figure~\ref{fig:representative}; right). All other interface features, task materials, time limits, opportunities to request feedback, and prompts to the LLM (except exclusion of trace) were held constant across conditions.

\subsubsection{Study Procedure}
The study (approved by IRB) consisted of four main stages: (1) guided practice (up to 10 minutes), (2) plan-tracing (P condition) or planning (N condition) (15 minutes), (3) coding (10 minutes), and (4) reflective semi-structured interview (up to 10 minutes). After consent and demographics, participants viewed a worked example~\cite{worked_examples}  that demonstrated the tool used in their assigned condition and completed a short practice task.
They then worked on the study task (\textit{Jump Game} problem from LeetCode\footnote{\url{https://leetcode.com/problems/jump-game/description/}}~\cite{dbox}), iteratively revising their plan using AI-generated feedback, then completed the coding task in Python without AI assistance while referencing their plan. 
Participants were asked to think aloud during plan-tracing (or planning only) and coding. 
Then, they completed a survey with 7-point Likert-scale items measuring perceived learning, confidence, intentions to use a similar workflow in the future, cognitive load adapted from ~\cite{leppink2013development}, and interface usability using UMUX-LITE~\cite{umux-lite}. 
The session concluded with a semi-structured interview about participants' planning and coding processes, their experiences with AI feedback, and perceived benefits and limitations of the workflow.

\subsection{Data Analysis}

\subsubsection{Data Collection}
We collected four primary sources of data: (1) plan submissions (and the associated traces in the plan-tracing condition) and the corresponding LLM-generated feedback, (2) participants' final code submissions and their performance on automated test cases, (3) responses to the demographic and post-task surveys, and (4) think-aloud and post-task interview transcripts.

\subsubsection{Plan Representation Analyses} 
\label{sec:plan-rep}
We conducted a qualitative analysis of students’ plans. Following a hybrid inductive and deductive approach~\cite{clarke2017thematic,fereday2006demonstrating}, the first author inductively examined the final plans to identify recurring representational patterns and develop a codebook. Two authors then independently applied the codebook to all 20 final plans and resolved disagreements through discussion (mean Cohen's $\kappa$ across all coded dimensions = 0.82, suggesting high agreement \cite{landis1977measurement}). We complemented this with a quantitative exploration. For each plan submission, we measured (1) the number of steps, (2) word count by splitting the plan text on whitespace, and (3) the number of control-flow references, identified using keywords such as ``if'', ``elif'', ``else'', ``while'', and ``return''. To account for participants submitting multiple plans, we modeled each outcome using a mixed-effects regression with condition and submission order as fixed effects and participant as a random intercept.

\subsubsection{Code Performance and Survey Analyses}
We evaluated students’ code performance by recording the number of LeetCode test cases passed. We compared conditions on code performance and survey measures using two-sided Mann--Whitney U tests.

\subsubsection{LLM-Generated Feedback Analyses}
We evaluated LLM feedback on two dimensions: correctness in identification of the earliest erroneous plan step (0 = incorrect, 1 = correct) and explanation quality (0 = incorrect, 0.5 = correct but vague, 1 = correct and informative enough to support revision). Two coders independently rated 15\% of all responses and resolved disagreements through discussion (mean percent agreement = 67\%\footnote{Because the code distributions were highly imbalanced, Cohen's $\kappa$ would be difficult to interpret. We therefore report observed percent agreement.}). One coder then rated the remaining responses. As in Section~\ref{sec:plan-rep}, we analyzed feedback correctness using a logistic mixed-effects regression and explanation quality using a linear mixed-effects regression.

\subsubsection{Think-aloud and Semi-structured Interview Analysis.}
We conducted a thematic analysis of the full interview transcripts to understand participants’ experiences with the workflow. 

\label{sec:methods}

\section{Results}
\subsection{How Tracing Shaped Planning}
\label{sec:results_rq1}
We designed plan tracing to help students articulate their approach more precisely, so we first explored differences in how students represented their plans between conditions. 

\subsubsection{Tracing shifted plans from code-like operations to goal-level descriptions.}
\label{sec:results_rq1_code_goals}
Students in the plan-tracing condition often compressed multiple operations into a single goal-level step, whereas those in the planning-only condition tended to separate actions into finer-grained steps that more closely mirrored code structure. 
T09\footnote{Throughout the results, we use ``T'' for quotes from participants in the plan-tracing condition and ``N'' for those in the planning-only condition.}, for example, expressed both the search for the maximum reachable position and the subsequent pointer movement as one high-level goal-level step: \planstep{``Find the location that gives the maximum possible distance that can be reached and jump to that location''}. 
In contrast, N07 expressed the same operation as two separate, more concrete code-like steps: \planstep{``set tracker as the nums[counter] + counter''} and \planstep{``set counter as tracker and loop back up''}. 

Even quantitatively, plans in the plan-tracing condition contained fewer steps ($\beta=-2.03$, $SE=0.91$, $p=.026$) and control-flow references ($\beta=-3.51$, $SE=1.48$, $p=.018$) than those in the planning-only condition, but did not differ in total word count. These results together support the qualitative observation that students in the plan-tracing condition put together multiple operations into fewer, higher-level steps rather than simply writing less.

\subsubsection{Tracing encouraged students to articulate rationale, not only actions.}
\label{sec:results_rq1_rationale}

Beyond structure, plans from the plan-tracing condition more often articulated the intent behind operations, whereas the plans from the planning-only condition tended to list actions without rationale. T04, for instance, explicitly connected operations to their purpose, \planstep{``The furthest position will be the largest number within nums [...] I want to get that maximum and add that to tracker to move forward, and then use that to iterate counter forward''}. N03, by contrast, described only the logic in prose, \planstep{``if counter greater than or equal to length of num return false''}. Students’ reflections help explain the difference. T05, whom we observed constructing the trace before writing their natural-language description, explained that ``sometimes I, myself, don't really know what I'm trying to do, so it's easier to just mess around with the numbers before converting those abstract thoughts into more plain language'' and described tracing as a way to ``explain [the algorithm] to yourself.'' For these students, tracing seemed to help them think and construct solutions that they narrated as the rationale for the plans.

\subsection{How Plan Tracing Affected AI Feedback}
\label{sec:results_rq2}

We designed plan tracing to give the LLM clear evidence of the plan's intended behavior, reducing its reliance on inferring intent from natural language alone leading to generation of more accurate feedback (see Section~\ref{sec:system}). We therefore examined the feedback students received and how they experienced it.

\subsubsection{AI feedback was understandable and useful for validation, but often insufficient for revision.}
\label{sec:results_rq2_feedback_errors}

Students in both conditions identified the AI guidance as one of the most useful features of the tool. When they trusted the feedback, they expressed that it helped validate their reasoning, making them more confident in the code writing task (T06, T09). 
Despite this usefulness, students under both conditions reported three recurring failure modes in the LLM feedback: \textit{insufficient explanation}, \textit{inconsistent feedback across iterations}, and \textit{limited representational flexibility}. 

First, participants found the feedback insufficient, but not unclear. T05 explained that ``the feedback wasn't confusing itself, it just wasn't enough for, like, how less I knew,'' and others noted that the feedback identified errors without explaining why their reasoning was wrong or how to meaningfully revise it (T05, T06, T08, N01). Second, participants reported reduced trust when the system showed inconsistent responses across iterations. Some described that the system contradicted itself by flagging steps it had previously marked as correct. T09 highlighted that, ``I wrote one of the lines, the LLM said it was correct. I went to the second line, I changed something, the LLM didn't say anything about the line I changed, but it said the line I changed before was incorrect.'' Others reported the system repeating the same feedback even after the flagged step has been revised (N09, N07). Third, participants felt the LLM feedback limited how they could represent their approach. Although the tool was designed to allow flexibility in the level of detail, participants felt pushed toward more verbose, highly detailed descriptions (T07, N04, N05). Some felt the fixed number and type of variables were too restrictive and forced them toward solutions that did not match their intended approach (T02, T03). These failure modes were reported by participants in both conditions, suggesting that tracing did not change how students experienced the feedback. 

\subsubsection{Feedback quality was similar across conditions.}
\label{sec:results_rq1_feedback_qual}

We qualitatively analyzed whether the LLM correctly localized the first erroneous step in each plan submission and rated the quality of the feedback (Section ~\ref{sec:methods}). The LLM correctly localized the first erroneous step in 74\% of plan-tracing submissions and 80\% of planning-only submissions, with mean explanation quality of $0.66$ and $0.69$, respectively. The mixed-effects models did not yield significant results, suggesting that feedback was similar across conditions and echoing students' own experiences that the trace did not meaningfully change the feedback they received.

\subsection{How Students Perceived Plan Tracing}

\label{sec:results_rq3}
\subsubsection{Tracing helped students verify and refine their plans before coding.}
\label{sec:results_rq3_verify}

Students described tracing primarily as a mechanism for verifying their plans before writing any code. They used it to identify emerging patterns in the problem to construct a solution (T08), iteratively test small examples (T01, T05), surface edge cases (T02, T05, T07), and revise their approach (T01, T05). T03 noted that plan tracing could nudge students to test their plan on multiple inputs, which then could translate into greater confidence during implementation. T05 also described that ``[...] it makes it [coding task] a lot easier and a little less daunting [...] if, like, you write down [...] iterations, like, before you actually look at the IDE.''  Overall, nine of ten participants expressed finding the trace-based planning valuable, describing it as beneficial for how they approached and practiced problem-solving rather than for learning new domain concepts (T05). Several volunteered uses beyond the study task, such as preparing for technical interviews by practicing how to clearly describe, explain, and validate an approach (T04).

\subsubsection{Tracing was most useful when students were uncertain, but tedious when the solution felt obvious.}
\label{sec:results_rq3_tedious}

Tracing's value, however, was conditional. T08 noted that on larger, more complex inputs ``it would be really tedious to set tracker and counter to whatever the number should be,'' and ``if I really didn't know what was going on, then [tracing] would be really helpful, because I could track through each step [...] but when I kind of understand it, [tracing] would be too tedious''. This suggests that students may find tracing most worthwhile when they are uncertain of their approach and when the input is small enough to trace by hand.

\subsubsection{Students wanted to revisit and revise plans during implementation.}
\label{sec:results_rq3_linear}
Other experiences of using plans for coding were shared across conditions. A plan's value as a reference depended on whether participants believed it was correct (T05, N04). Participants in both conditions found that writing code surfaced logical issues that planning alone did not. N01 reported an ``epiphany when writing the code'' that reshaped their approach, and T01 said that their ``thought process defogs while coding,'' with corner cases emerging only at implementation time. Students consequently wanted to revisit their earlier work as these issues arose; T01 and T05 wanted to update their trace during coding to simulate newly discovered corner cases and T05 and N03 wanted to change their overall approach mid-implementation and receive AI feedback on the revised strategy.

\subsubsection{Coding performance was comparable across conditions, but confidence was lower in plan-tracing.}
\label{sec:results_rq3_performance}

Out of 178 test cases on LeetCode, participants from the planning-only condition passed 79.00 on average (MD = 44.00, SD = 86.27) and those from the plan-tracing condition passed 60.90 (MD = 39.50, SD = 67.21), however, these differences were not significant. Despite comparable performance, participants in the planning-only condition reported higher confidence in their skills to solve similar problems in the future ($U = 21.5$, $p = .025$, $r = -.50$). Some participants in the plan tracing condition (T05, T08) attributed this uncertainty to the concept rather than the workflow: ``I feel confident with tracing, but maybe not specifically greedy algorithms'' (T05). Another explanation is that tracing surfaced edge cases and flaws in students’ solutions (Section~\ref{sec:results_rq3_verify}). Confronting these newly visible problems may have made the task feel more difficult and reduced participants’ sense of mastery~\cite{deslauriers_illusion_of_learning}. We observed no differences between conditions in perceived learning, workflow adoption, or usability.

\label{sec:results}

\section{Discussion and Conclusion}
\subsection{Tracing Changed Plans but Not AI Feedback}

Plan-tracing changed how students expressed their plans. We expected plan tracing to help learners describe their intent more precisely (Section~\ref{sec:system}). In contrast, plans shifted away from describing concrete execution details toward higher-level, goal-driven steps. Students constructed plans with fewer steps and combined multiple operations within a single step (Section~\ref{sec:results_rq1_code_goals}), suggesting they organized individual actions into higher-level subgoals~\cite{margulieux2012subgoal}. In addition, these plans often described \textit{what} a step should accomplish and \textit{why} it was needed, instead of focusing on \textit{how} it should be implemented (Section~\ref{sec:results_rq1_rationale}). One explanation is that students used the descriptions to express goals and rationale, letting the trace demonstrate the specific execution behavior on an input.

We also designed plan tracing to improve the quality of AI-generated feedback (Section~\ref{sec:system}). However, we did not observe any meaningful differences in the quality of the feedback between conditions (Section~\ref{sec:results_rq2}). The shift described above may help explain why. In the planning-only condition, the LLM inferred students’ intent from code-like steps written in natural language. In the plan-tracing condition, the trace captured execution behavior, while the plan conveyed goals and rationale. The code-like steps and the execution trace may have provided similarly concrete evidence of the algorithm’s behavior, allowing the model to infer students’ intent to a similar degree and provide comparable feedback.

\subsection{Implications for Instructors and Tool Designers}

Our observations (Section~\ref{sec:results_rq2} and ~\ref{sec:results_rq3}) surfaced several trade-offs in how such tools may be developed and used in instruction.

\subsubsection{For Instructors} 
Our workflow required all students in the plan-tracing condition to trace their written algorithm on the given input before coding (Section~\ref{sec:system_coding}). In practice, however, students wanted to use tracing selectively. They found it most useful when they were uncertain about their approach, but tedious when the solution seemed obvious (Section~\ref{sec:results_rq3_tedious}). We also designed the AI feedback to be fix-withholding, which is identifying the error and where it occurred without revealing how to fix it (Section~\ref{sec:system_feedback}). This preserved students’ responsibility for revision, but some learners found the feedback too limited to be helpful (Section~\ref{sec:results_rq2_feedback_errors}). Both design decisions, therefore, imposed a single level of support that served experienced and novice learners unevenly. Instructors could:

\begin{instructorbox} Adapt planning scaffolds and AI feedback to learner confidence and expertise, providing more support for novices and fading guidance \cite{atkinsonfadedscaffolding} as independence grows. \end{instructorbox}

We designed the planning-tracing-coding workflow as a linear sequence where students first developed a correct plan that they could then use as a solution to translate into code (Section~\ref{sec:system_coding}). However, students expressed that implementation revealed edge cases that planning alone did not. Consequently, some students wanted to revise their plan or trace during the implementation stage (Section~\ref{sec:results_rq3_linear}). Instructors could:

\begin{instructorbox} Create an integrated workflow where planning, tracing, and coding inform each other throughout the problem-solving process, rather than as steps in a fixed linear sequence.
\end{instructorbox}

\subsubsection{For Tool Designers} 
We required students to use a fixed set of variables so the system could compare their traces with an expert trace and provide reliable feedback (Section~\ref{sec:system_feedback}). However, students found this restrictive because it pushed them to approaches they had not originally intended (Section~\ref{sec:results_rq2_feedback_errors}). This reveals a trade-off between reliable feedback and representational flexibility where comparing against a reference makes feedback reliable, but narrows the range of approaches a student can use. Tool designers could:

\begin{designerbox} Explore strategies that balance reliable evaluation with learner flexibility, surfacing flaws within a student's own approach rather than steering them toward canonical solutions.
\end{designerbox}

Participants reported reduced trust in the AI feedback when it was inconsistent across iterations, such as when the system contradicted its earlier evaluations or repeated the same message after the step had been revised (Section~\ref{sec:results_rq2_feedback_errors}). Students were unsure whether their revision introduced a new issue or whether the feedback was unreliable. Because the feedback was regenerated independently at each verification, it carried no memory of previous submissions or what the student had changed between them. Tool designers could:
\begin{designerbox} Incorporate memory of prior submissions into the prompting pipeline to provide consistent feedback across iterations. \end{designerbox}

\subsection{Broader Applications of Plan Tracing}
Beyond the specific tool developed in this study, plan-tracing can be integrated into existing activities and assessments as a problem-solving technique that students learn and practice when solving programming problems. It may be especially useful for teaching Prompt Problems~\cite{denny2024prompt}, where students specify programs in natural language rather than writing code. By tracing their plans before implementation, students can learn to verify their intended algorithm before delegating code generation to an LLM. More broadly, as natural language becomes a common medium for programming with GenAI, mismatches between learners’ intentions and authored  specifications become more consequential. Plan tracing offers one concrete way to make those mismatches visible and help learners resolve them.

\label{sec:discussion}

\begin{acks}
We thank our anonymous reviewers. 
\end{acks}


\bibliographystyle{ACM-Reference-Format}
\bibliography{sample-base}

\appendix

\end{document}